\documentclass[aps,pre,reprint,groupedaddress]{revtex4-1}

\usepackage{amsmath,amssymb,amsfonts}
\usepackage{bm}
\usepackage{hyperref}
\usepackage{epsfig,graphicx}
\usepackage{cancel}
\usepackage{color}
\usepackage[utf8]{inputenc}

\newcommand{\ra}{\rangle}
\newcommand{\la}{\langle}

\newcommand{\be}{\begin{equation}}
\newcommand{\ee}{\end{equation}}

\newcommand{\vv}{\bm{v}}
\newcommand{\rr}{\bm{r}}

\newcommand{\eq}{\text{eq}}
\newcommand{\st}{\text{st}}
\newcommand{\ini}{\text{ini}}
\newcommand{\diff}{\text{diff}}
\newcommand{\noise}{\text{noise}}
\newcommand{\inel}{\text{inel}}

\newcommand{\lin}{\text{lin}}

\newcommand{\tder}[1]{\partial_{t}{#1}}
\newcommand{\xder}[1]{\partial_{x}{#1}}
\newcommand{\vder}[1]{\partial_{v}{#1}}
\newcommand{\xxder}[1]{\partial_{x}^{2}{#1}}
\newcommand{\vvder}[1]{\partial_{v}^{2}{#1}}

\newcommand{\calH}{\mathcal{H}}

\begin{document}


\title{Global stability and $H$-theorem in lattice models
  with non-conservative interactions}



\author{C.~A.~Plata}
\email[]{cplata1@us.es}
\author{A.~Prados}
\email[]{prados@us.es}
\affiliation{F\'{\i}sica Te\'orica, Universidad de Sevilla, Apdo.\ de Correos 1065, Sevilla 41080, Spain}


\date{\today}

\begin{abstract}
  In kinetic theory, a system is usually described by its one-particle
  distribution function $f(\rr,\vv,t)$, such that
  $f(\rr,\vv,t)d\rr d\vv$ is the fraction of particles with positions
  and velocities in the intervals $(\rr,\rr+d\rr)$ and
  $(\vv,\vv+d\vv)$, respectively. Therein, global stability and the
  possible existence of an associated Lyapunov function or $H$-theorem
  are open problems when non-conservative interactions are present, as
  in granular fluids.  Here, we address this issue in the framework of
  a lattice model for granular-like velocity fields. For a quite
  general driving mechanism, including both boundary and bulk driving,
  we show that the steady state reached by the system in the long time
  limit is globally stable. This is done by proving analytically that
  a certain $H$-functional is non-increasing in the long time
  limit. Moreover, for a quite general energy injection mechanism, we
  are able to demonstrate that the proposed $H$-functional is
  non-increasing for all times. Also, we put forward a proof that
  clearly illustrates why the ``classical'' Boltzmann functional
  $H_{B}[f]=\int\! d\rr \, d\vv f(\rr,\vv,t) \ln f(\rr,\vv,t)$ is
  inadequate for systems with non-conservative interactions. Not only
  is this done for the simplified kinetic description that holds in
  the lattice models analysed here but also for a general kinetic
  equation, like Boltzmann's or Enskog's.
\end{abstract}


\maketitle


\section{Introduction}
\label{intro}

In thermodynamics and statistical mechanics, global stability of the
equilibrium state is usually proven by introducing a Lyapunov
functional \cite{lyapunov_general_1992}. This Lyapunov functional of
the probability distribution function (PDF) has the following three
properties: (i) it is bounded from below, (ii) it monotonically
decreases with time and (iii) its time derivative equals zero only
when the PDF is the equilibrium one. Therefore, in the long time
limit, the Lyapunov functional must tend to a finite value and thus
its time derivative vanishes. As a consequence, any PDF, corresponding
to an arbitrary initial preparation, tends to the equilibrium PDF: the
equilibrium state is irreversibly approached and said to be globally
stable.

The first example of such a Lyapunov functional is the renowned
Boltzmann $H$-functional. In the Boltzmann description, the
nonequilibrium behaviour of a dilute gas is completely encoded in the
one-particle velocity distribution function $f(\rr,\vv,t)$. By
introducing the Stosszahlansatz or Molecular Chaos hypothesis,
Boltzmann derived a closed non-linear integro-differential equation
for $f(\rr,\vv,t)$ governing its time evolution \cite{B95}. Also, for
a spatially homogeneous state, he showed that the functional
$H_{B}[f]=\int d\vv f(\vv,t) \ln\! f(\vv,t)$ has the three properties
of a Lyapunov functional. This $H$-theorem shows that all solutions of
the Boltzmann equation tend in the long time limit to the Maxwell
velocity distribution and irreversibility naturally stems from a
molecular picture
\cite{lebowitz_boltzmanns_1993,lebowitz_macroscopic_1993}. Interestingly,
a key point for deriving the $H$-theorem is the reversibility of the
underlying microscopic dynamics. In an inhomogeneous situation, one
has to consider the spatial dependence of the one-particle
distribution function $f(\rr,\vv,t)$, and the above functional must be
generalised to
\begin{equation}\label{H-Boltzmann}
H_{B}[f]=\int d\rr \, d\vv f(\rr,\vv,t) \ln\! f(\rr,\vv,t).
\end{equation}
Provided that the walls of the gas container are smooth, in the sense
that there is no energy transport through them, it can be also shown
that this is a non-increasing Lyapunov functional \cite{CC70}.

Another example of a Lyapunov functional can be found in the realm of
Markovian stochastic processes. Therein, the stochastic process $X(t)$
is completely determined by the conditional probability density
$P_{1|1}(X,t|X_{0},t_{0})$ of finding the system in state $X$ at time
$t$, given it was in state $X_{0}$ at time $t_{0}$, and the
probability density $P(X,t)$ of finding the system in state $X$ at
time $t$ \cite{van_kampen_stochastic_1992}. Both probability densities
satisfy the same evolution equation, named the master equation, but
with different initial conditions: one always has that
$P_{1|1}(X,t_{0}|X_{0},t_{0})=\delta(X-X_{0})$, whereas
$P(X,t_{0})=P_{\ini}(X)$, with $P_{\ini}(X)$ corresponding to the
(arbitrary) initial preparation. When the stochastic process is
irreducible or ergodic, that is, every state can be reached from any
other state by a chain of transitions with non-zero probability, there
is only one stationary solution of the master equation. In physical
systems, this steady solution must correspond to the
equilibrium-statistical-mechanics distribution $P_{\eq}(X)$. What is
more, a Lyapunov functional can be constructed in the following way,
\begin{equation}\label{H-function-master-eq}
  \calH[P]=\int dX P_{\text{eq}}(X) \, g\!\left[\frac{P(X,t)}{P_{\eq}(X)}\right],
\end{equation}
where $g(x)$ is any positive-definite convex function
($g''(x)\geq 0$). It must be stressed that the proof of this
$\calH$-theorem for master equations rely only on the ergodicity of
the underlying microscopic dynamics: it is not necessary to assume that
detailed balance, which is connected with the microscopic
reversibility,  holds \cite{van_kampen_stochastic_1992}.

The most usual choice for $g$ is $g(x)=x\ln x-x+1$, which leads to
\begin{equation}\label{H-function-add}
\calH[P]=\int dX P(X,t)\, \ln\!\left[\frac{P(X,t)}{P_{\eq}(X)}\right].
\end{equation}
The physical reason behind this choice is the ``extensiveness'' of
$\calH[P]$: if the system at hand comprises two independent subsystems
$A$ and $B$, so that $dX\equiv dX_{A}dX_{B}$ and
$P(X)=P_{A}(X_{A})P_{B}(X_{B})$, one has that
$\calH[P]=\calH_{A}[P_{A}]+\calH_{B}[P_{B}]$. It is to consider
$-\calH$ as a nonequilibrium entropy $S$ that this extensiveness is
desirable: in this way, the non-increasing behaviour of $\calH$ leads
to a non-decreasing time evolution of $S$. Moreover, $H[P]$ remains
invariant upon a change of variables $Y=f(X)$, as emphasised in Refs.~\cite{marconi_about_2013,de_soria_towards_2015}.

Although the Boltzmann equation is not a master equation, we may
wonder why the expressions for $H_{B}$ in Eq.~\eqref{H-Boltzmann}
and $\calH[P]$ in Eq.~\eqref{H-function-add} are
different. Specifically, we may wonder why not writing
\begin{equation}\label{H-one-particle}
H[f]=\int d\rr \, d\vv f(\rr,\vv,t) \ln\! \left[\frac{f(\rr,\vv,t)}{f_{\eq}(\vv)}\right]
\end{equation}
instead of $H_{B}[f]$. Up to now, we have been implicitly considering
the ``classic'' problem with elastic collisions between particles, in
which the system eventually reaches thermodynamic
equilibrium. Therein, the answer is trivial: since $\ln f_{\eq}(\vv)$
is a sum of constants of motion, $H[f]-H_{B}[f]$ is constant and both
are utterly equivalent.

Whether there exists an extensive $H$-functional or not is an
important question in nonequilibrium statistical physics. If the
answer were positive, it would make it possible to define a
non-equilibrium entropy $-H$ that monotonically grows for all times,
extending the Clausius inequality. In general, the system at hand does
not reach equilibrium but a nonequilibrium steady state. Thus, the
equilibrium distribution $f_{\eq}$ in $H$ has to be substituted with
the stationary one $f_{\st}$. In this context, the field of granular
fluids is a benchmark for intrinsically out-of-equilibrium,
dissipative, systems: the microscopic dynamics is not time-reversible
because collisions between particles are inelastic, but a
nonequilibrium steady state can be attained if some driving mechanism
injects energy into the system.

In granular fluids, the functionals $H[f]$ and $H_{B}[f]$ are no
longer equivalent, since $\ln f_{\st}$ is not a sum of constants of
motion. Indeed, for granular gases described by the inelastic
Boltzmann equation \cite{PL01,Vi06}, there are some results that hint at
$H_{B}$ not being a Lyapunov functional. Within the first
Sonine approximation, it has been proven that the time derivative of
$H_{B}$ does not have a definite sign in the linear approximation
around the steady state \cite{bena_stationary_2006}. Moreover, Marconi
et al.~have numerically shown that $H_{B}$ is non-monotonic and
even steadily increases from certain initial conditions
\cite{marconi_about_2013}. They have also put forward some
numerical evidence (further reinforced by Garc\'ia de Soria et
al.~\cite{de_soria_towards_2015}) in favour of $H$ being a ``good''
Lyapunov functional. Notwithstanding, only spatially homogeneous
situations, in which the $\rr$-dependence of $f$ and thus the
integration over $\rr$ may be dropped, have been analysed in 
Refs.~\cite{marconi_about_2013,de_soria_towards_2015}. 

Some years ago, a simplified model for a granular-like velocity field
was introduced to study correlations in granular gases
\cite{BMP02}. Very recently, a variant of this model on a
one-dimensional lattice has been proposed to mimic the velocity
component along the shear direction \cite{lasanta_fluctuating_2015},
and both its hydrodynamic limit and finite size effects have been
analysed
\cite{lasanta_fluctuating_2015,manacorda_lattice_2016,plata_lattice_2016}. This
model has been shown to retain a relevant part of the granular
phenomenology: the shear instability of the homogeneous cooling state,
the existence of boundary driven steady states such as the Couette and
Uniform Shear Flow (USF) states, the renormalisation of the cooling
rate due to fluctuations close to the shear instability, etc. Other
properties thereof, when it is driven by a mechanism resembling
collisions with a randomly moving inelastic wall, have been studied in
Ref.~\cite{prasad_driven_2016}. At the $N$-particle level, the
dynamics of the system is governed by a master equation, which is
analogous to the Kac equation \cite{Ka56}, that leads to a ``kinetic''
equation at the one-particle level, which is analogous to the
Boltzmann equation. In the latter, the collision term, although being
simpler than that in the Boltzmann equation, remains a non-linear
integro-differential one \cite{manacorda_lattice_2016}.

It must be recalled that an analytical proof of either global
stability or the $H$-theorem is currently unavailable at the level of
the kinetic description for granular gases. This is true even for
simple collision terms, such as those corresponding to hard-spheres or
the cruder Maxwell particle model (where the collision rate is
considered to be velocity-independent), which are considered in the
pioneering work in
Refs.~\cite{marconi_about_2013,de_soria_towards_2015}. Therefore, it
seems worth investigating this subject in simplified models, for which
analytical calculations are more feasible.

Our main goal here is to investigate the global stability and the
possibly associated $H$-theorem in the above class of lattice
models. Unlike the approach in
Refs.~\cite{marconi_about_2013,de_soria_towards_2015}, we do not
restrict ourselves to spatially homogeneous situations but consider
the whole space and velocity dependence of the one-particle PDF
$f(\rr,\vv,t)$. Specifically, we introduce a general energy injection
mechanism, in which the system may be driven both through the
boundaries and in the bulk. We show that, under quite general
conditions, the steady state is globally stable: independently of the
initial preparation, the system always ends up in the steady
state. Interestingly, it is not necessary to have an $H$-theorem to
prove this: it suffices to show that $H$ is decreasing in the long
time limit, not for all times. In this sense, the situation is
analogous to the proof of the tendency towards the equilibrium curve
in systems whose dynamics is governed by master equations with
time-dependent transition rates
\cite{brey_normal_1993,brey_dynamical_1994,brey_dynamical_1994-1,vlad_deterministic_1997,vlad_h-theorem_1998,prados_hysteresis_2000,earnshaw_global_2010}.

Our proof of global stability also enables us to show the inadequacy
of Boltzmann's $H_{B}$ as a candidate for Lyapunov functional in
inelastic systems. Not only is this done for the simplified models
considered in the paper, but for a general collision term that does
not conserve energy in collisions. Therefore, this result also applies
to the inelastic Boltzmann or Enskog equations used in granular
fluids. The main idea is that the sign of $dH_{B}/dt$ can be reversed
by a suitable choice of the initial PDF, and thus cannot have a
definite sign. In this respect, our result generalises that in
Ref.~\cite{bena_stationary_2006}, which was derived within the first
Sonine approximation of the inelastic Boltzmann equation, to an
arbitrary collision kernel with non-conservative interactions.

Having proved global stability by showing that $H$ is a non-increasing
functional for long times, a natural question remains. Is it $H$ a
Lyapunov function, that is, a non-increasing functional for all times?
There does not seem to be a unique proof, valid for any driving
mechanism, even within our simplified model. Nevertheless, we have
been able to derive a specific proof for a quite general driving
mechanism, which includes as limiting cases both the sheared system,
in which the steady state is the USF state, and the uniformly heated
system by means of the so-called stochastic thermostat
\cite{van_noije_velocity_1998,MS00,GMyT09,GMT12,PTNE02,maynar_fluctuating_2009,prados_kovacs-like_2014,trizac_memory_2014,prasad_high-energy_2013}. The
proof is based on a suitable expansion of the one-particle PDF in
Hermite polynomials, which is a generalisation of the usual Sonine
expansion of kinetic theory.

The paper is organised as follows. In Sec.~\ref{model}, we briefly
introduce the model, its dynamics and the continuum limit. Section
\ref{global-stability} is devoted to the proof of the global stability
of the nonequilibrium steady states, for a general energy injection
mechanism. The inadequacy of Boltzmann's $H_{B}$ as a Lyapunov
functional for inelastic systems is discussed in
Sec.~\ref{sec:inadequacy}.  Later, in Sec.~\ref{USF}, we consider some
concrete physical situations in our model, which include the sheared
and the uniformly heated systems. Therein, we show that $H[f]$ is a
monotonically decreasing Lyapunov functional.  Finally,
Sec.~\ref{conc} gives the main conclusions of the paper. Some
technical details, which are omitted in the main text, are given in
the Appendices.

\section{The model: dynamics and continuum limit}
\label{model}

Here, we present the general class of models that was introduced in
Ref.~\cite{lasanta_fluctuating_2015}, focusing on the continuum description
obtained in the large system size limit
\cite{manacorda_lattice_2016}. Specifically, our system is defined on
a $1$d lattice: at each lattice site $l$, there is a particle with
velocity $v_{l}$. Thus, at a given time $\tau$, the configuration of
the system is completely determined by
$\vv \equiv \{v_{1},...,v_{N}\}$. The dynamics proceeds
through inelastic nearest-neighbour  binary collisions: each pair
$(l,l+1)$ collides inelastically with a characteristic rate
$\omega^{-1}$, independently of their relative velocity (the so-called
Maxwell-molecule model \cite{BK03}) and the state of the other
pairs. We introduce the operator $\hat{b}_{l}$ that transforms the
pre-collisional velocities into the post-collisional ones,
\begin{subequations}\label{eq:coll_rule}
\begin{eqnarray}
\hat{b}_{l}v_{l} &= v_{l}-\frac{1+\alpha}{2}\left(v_{l}-v_{l+1}\right), \\
\hat{b}_{l}v_{l+1} &= v_{l+1}+\frac{1+\alpha}{2}\left(v_{l}-v_{l+1}\right),
\end{eqnarray}
\end{subequations}
where $\alpha$ is the normal restitution coefficient, with
$0<\alpha\leq 1$. 

In addition to collisions, the system is heated by a stochastic force
that is modelled by a white noise, the so-called \textit{stochastic
  thermostat}
\cite{van_noije_velocity_1998,MS00,GMyT09,GMT12,PTNE02,maynar_fluctuating_2009,prados_kovacs-like_2014,trizac_memory_2014,prasad_high-energy_2013}. Specifically,
for a short time interval, the change of the velocity due to the
heating is given by
\begin{eqnarray}
  \left.\Delta v_{i}(\tau)\right|_{\text{noise}}&\equiv &\left. v_{i}(\tau+\Delta
                                                       \tau)-v_{i}(\tau)\right|_{\text{noise}} \nonumber \\ &=&\left(\xi_{i}(\tau)-\frac{1}{N}\sum_{j=1}^{N}
                                                                                                          \xi_{j}(\tau)\right)\Delta \tau, \label{jump-moments-1}
\end{eqnarray}
where $\xi_{i}(t)$ are Gaussian white noises, verifying
\begin{equation}\label{jump-moments-2}
\la \xi_{i}(\tau)\ra_{\text{noise}}=0, \quad \la \xi_{i}(\tau)\xi_{j}(\tau')\ra_{\text{noise}}=\chi
\delta_{ij}\delta(\tau-\tau'),
\end{equation}
for $i,j=1,\ldots,N$. Above, $\chi$ is the amplitude of the noise, and
$\la\cdots\ra_{\text{noise}}$ denotes the average over the different
realisations of the noise. Note that this version of the stochastic
thermostat conserves total momentum, a necessary condition to have a
steady state \cite{maynar_fluctuating_2009,prasad_high-energy_2013}.

We define $P_{N}(\vv,\tau)$ as the probability density of finding the
system in state $\vv$ at time $\tau$. The stochastic process
$\vv(\tau)$ is Markovian and the equation governing the time evolution
of $P_{N}(\vv,\tau)$ has two
contributions. First, we have a master equation contribution stemming
from collisions
 \cite{manacorda_lattice_2016,plata_lattice_2016}
\begin{equation} \label{eq:master-equation}
\left.\partial_{\tau} P_N(\vv,\tau)\right|_{\text{coll}}=\omega \sum_{l=1}^N \left[ \frac{P_{N}(\hat{b}_l^{-1} \vv,\tau)}{\alpha}  -  P_N(\vv,\tau) \right],
\end{equation}
in which the operator $\hat{b}_l^{-1}$ is the inverse of
$\hat{b}_{l}$, that is, it changes the post-collisional velocities
into the pre-collisional ones when the colliding pair is $(l,l+1)$. Second, there is a Fokker-Planck
contribution stemming from the stochastic forcing \cite{marconi_about_2013,de_soria_towards_2015}
\footnote{This kind of Fokker-Planck kernel is similar to the one appearing in
the Fokker-Planck equation for biomolecules in which their total
length $L=\sum_{l=1}^{N}\eta_{l}$, where $\eta_{l}$ is the length of
each of the modules of the molecule, is controlled and kept
constant. This is quite logical, since in both situations there is an
additive conservation law: here it is the total momentum
$\sum_{l=1}^{N}v_{l}$ that is conserved, whereas in biomolecules the
constant of motion is the total length \cite{bonilla_theory_2015}.}
\begin{equation} \label{eq:Fokker-Planck}
\left.\partial_\tau
  P_N(\vv,\tau)\right|_{\text{noise}}=\frac{\chi}{2}
\sum_{i,j=1}^{N}\left(\delta_{ij}-\frac{1}{N}\right)\frac{\partial^{2}}{\partial
v_{i}\partial v_{j}}P_{N}(\vv,\tau).
\end{equation}
The time evolution of $P_{N}(\vv,\tau)$ is obtained by combining
Eqs.~\eqref{eq:master-equation} and \eqref{eq:Fokker-Planck}, that is,
\begin{equation}\label{time-evol}
\partial_{\tau}P_{N}(\vv,\tau)=\left.\partial_{\tau}P_{N}(\vv,\tau)\right|_{\text{coll}}+\left.\partial_{\tau}P_{N}(\vv,\tau)\right|_{\text{noise}}.
\end{equation}

In this work, we focus on the evolution of quantities that can be
written in terms of the one-particle
distribution function, namely 
\begin{equation}\label{eq:P1}
 P_{1}(v;l,\tau)=\int d\vv P_{N}(\vv,\tau) \delta(v_{l}-v).
\end{equation} 
All the one-site velocity moments can be calculated from
$P_{1}$,
\begin{equation}
\la v_{l}^{n}(\tau)\ra\equiv \int_{-\infty}^{+\infty}dv\, v^{n} P_{1}(v;l,\tau).
\end{equation}
The first two moments give the hydrodynamic fields: the average
velocity $u_{l}(\tau)$ and granular temperature
$T_{l}(\tau)$ \footnote{Note that the density of the model is fixed,
  there is no mass transport in the system.},  which are defined by the
relations
\begin{equation}\label{eq:ul-Tl}
u_{l}(\tau)\equiv \la v_{l}\ra, \qquad T_{l}(\tau)\equiv \la v_{l}^{2}(\tau)\ra-u_{l}^{2}(\tau).
\end{equation}
Here, we do not write the evolution equations on the lattice for
either $P_{1}$ or the hydrodynamic fields ($u$ and $T$), since they
are not necessary for our present purposes. The unforced case
($\chi=0$) can be found in Ref.~\cite{manacorda_lattice_2016}. However, we
would like to stress that the evolution equation for $P_{1}$ is not
closed, since the collision term involves the two-particle
distribution function $P_{2}(v,v';l,l+1,\tau)$. As usual in kinetic
theory, one can write a closed equation for $P_{1}$ after introducing
the Molecular Chaos assumption, that is,
$P_{2}(v,v';l,l+1,\tau)=P_{1}(v;l,\tau)P_{1}(v';l+1,\tau)+O(N^{-1})$. In
other words, one assumes that the correlations at different sites are
of the order of $N^{-1}$ and thus negligible in the large system size
limit.

The continuum limit of the model is introduced for large system size
$N\gg 1$, in which we expect the average velocity $u_{l}$ and
temperature $T_{l}$ to be smooth functions of space and time. This is
expressed mathematically by defining ``hydrodynamic'' continuous space
and time variables by $x=l/N$ and $t=\omega\tau/N^{2}$, respectively
\cite{manacorda_lattice_2016}. Note that $0\leq x\leq 1$ and $t\geq 0$.
In the continuum limit, the one-particle distribution function also
becomes a smooth function of $x$ and $\tau$,
$P_{1}(v;x,t)\equiv P_{1}(v;l=Nx,\tau=N^{2}t/\omega)$. 

From now on, we use the usual notation in kinetic theory
$f(x,v,t)\equiv P_{1}(v;x,t)$. The physical picture is
straightforward: $f(x,v,t)dx dv$ gives the fraction of the total
number of particles with positions between $x$ and $x+dx$ and
velocities between $v$ and $v+dv$. We have that
$\int_{-\infty}^{+\infty}dv\,f(x,v,t)=1$ for all $x$ and $t$, since
there is no mass transport in the system.  The time evolution of $f$
is governed by the non-linear integro-differential (pseudo-Boltzmann)
equation \cite{manacorda_lattice_2016}
\begin{eqnarray}
\partial_{t}f=\partial_{x}^{2}f+\frac{\nu}{2}\partial_{v}\left\{[v-u(x,t)]
   f\right\}+\frac{\xi}{2}\partial_{v}^{2}f,
\label{eq:P1-hydroMM}
\end{eqnarray}
where $u(x,t)$ is the local average velocity,
$\nu$ is the macroscopic dissipation coefficient and $\xi$ is the
macroscopic noise strength, which are respectively given by
\begin{equation}\label{eq:nu}
  \nu=(1-\alpha^{2})N^{2}, \qquad \xi=\frac{\chi N^{2}}{\omega}.
\end{equation}
This shows that the microscopic noise strength $\chi$ must scale as
$N^{-2}$ in order to have a finite contribution in the continuum
limit. Of course, for $\xi=0$, we recover the kinetic equation for the
case in which there is no stochastic forcing, see
Ref.~\cite{manacorda_lattice_2016}. The $N$-scaling of the macroscopic
dissipation coefficient $\nu$ is similar to that found in the
dissipative version of the Kipnis-Marchioro-Presutti
model~\cite{kipnis_heat_1982,prados_large_2011,prados_nonlinear_2012,HLyP13,lasanta_statistics_2016}.

The average velocity $u(x,t)$ and granular temperature $T(x,t)$ are
the continuum limit of $u_{l}$ and $T_{l}$ defined in
Eq.~\eqref{eq:ul-Tl},
\begin{equation}\label{hydro-fields-continuum}
u(x,t)\equiv \la v\ra(x,t), \qquad T(x,t)\equiv \la v^{2}\ra(x,t)-u^{2}(x,t),
\end{equation}
where the velocity moments are given by
$\la v^{n}\ra (x,t)=\int dv\, v^{n} f(x,v,t)$.  From the kinetic
equation for $f(x,v,t)$, one can derive the evolution equations of $u$
and $T$,
\begin{subequations}\label{eq:hydroMM}
\begin{align}
\partial_{t}u&=\partial_{xx} u, \label{eq:hydroMMu} \\
\partial_{t}T&=-\nu T+ \partial_{x}^{2} T+2\left(\partial_{x}u\right)^2+\xi . \label{eq:hydroMMT}
\end{align}
\end{subequations}
On the one hand, Eq.~\eqref{eq:hydroMMu} is a diffusion equation for
the average velocity, which expresses the conservation of total
momentum. On the other hand, the temperature equation
\eqref{eq:hydroMMT} contains a purely dissipative term $-\nu T$ that
stems from the inelasticity of collisions and always contributes to
``cooling'' the system, a diffusive term $\partial_{x}^{2}T$, a
viscous heating term $2\left(\partial_{x}u\right)^{2}$, and finally
the term corresponding to the uniform heating $\xi$. Of course, either
the kinetic equation for $f$ or the average equations for $(u,T)$ must
be complemented with suitable boundary conditions in each physical
situation.

\subsection{Non-equilibrium steady states and boundary conditions}\label{NESS}

We are interested in driven cases, in which there is an input of
energy that balances (in average) the energy loss in collisions, so
that the system eventually reaches a steady state. These
non-equilibrium steady states (NESS) are described by the corresponding
stationary solutions $f_{\st}(x,v)$ of the kinetic equation, which
verify
\begin{eqnarray}
0=\partial_{x}^{2}f_{\st}+\frac{\nu}{2}\partial_{v}\left\{[v-u_{\st}(x)]f_{\st}\right\} +\frac{\xi}{2}\partial_{v}^{2}f_{\st}.
\label{eq:P1-steady}
\end{eqnarray}
where $u_{\st}(x) = \int dv \, v f_{\st} (x,v)$ is the stationary
average velocity profile. To be concrete, we consider two cases: a
system that is (a) sheared and (b) uniformly heated.

First, let us consider a sheared system: there is no
stochastic forcing, $\xi=0$, and the driving is introduced by imposing
a velocity difference (``shear'') between the left and right edges of
the system. At the level of the hydrodynamic description, the
corresponding boundary conditions are
\begin{subequations}
  \label{eq:bc-USF_hydro}
  \begin{align}
    \label{eq:bc-USF-u}
    u(1,t)&=u(0,t)+a,  &u'(1,t)&=u'(0,t), \\
    \label{eq:bc-USF-T}
    T(1,t)&=T(0,t),   &T'(1,t)&=T'(0,t),
  \end{align}
\end{subequations}
which are said to be of Lees-Edwards type
\cite{LE72}. We have used $^{\prime}$ to denote spatial
derivative. The imposed shear allows the viscous heating term, which
is proportional to $(\partial_{x}u)^{2}$, to compensate for the energy
dissipation term, $-\nu T$.  The boundary conditions for the
one-particle distribution function read
\begin{subequations}\label{eq:P1-bc-USF}
\begin{align}  
 f(1,v,t)=f(0,v-a,t), \quad 
 f'(1,v,t)=f'(0,v-a,t),
\end{align}
\end{subequations}
from which Eq.~\eqref{eq:bc-USF_hydro} directly
follow. Equation \eqref{eq:P1-bc-USF} has a simple physical
interpretation: particles that leave the system through its right edge
with velocity $v$ are reinserted through its left edge with velocity
$v-a$.

The steady state for the sheared system is known as the USF state,
which has a linear velocity profile and a homogeneous temperature,
\begin{equation}\label{eq:ave-USF}
 u_{\st}(x)=a\left( x-\frac{1}{2} \right), \quad T_{\st}=\frac{2a^{2}}{\nu}.
\end{equation}
For our simplified model, the stationary PDF is Gaussian,
\begin{equation}\label{eq:P1-USF}
f_{\st}(x,v)=\left(2\pi T_{\st}\right)^{-1/2} \exp\left[-\frac{(v-u_{\st}(x))^{2}}{2T_{\st}}\right].
\end{equation}
An extensive investigation of the sheared system, at the level of the
average hydrodynamic equations, can be found in Ref.~\cite{manacorda_lattice_2016}.

Second, we address the uniformly heated system, in which there is no
shear, $a=0$, but there is stochastic forcing, $\xi\neq 0$. In this
case, we have the usual periodic boundary conditions. In particular,
for the PDF we have
\begin{equation}\label{eq:P1-bc-periodic}
 f(1,v,t)=f(0,v,t), \quad f'(1,v,t)=f'(0,v,t).
\end{equation}
In the steady state, the system is homogeneous: there is no average
velocity and the temperature is uniform,
\begin{equation}\label{eq:ave-Unif-heated}
 u_{\st}(x)=0, \quad T_{\st}=\frac{\xi}{\nu}.
\end{equation}
The corresponding stationary PDF is also Gaussian,
\begin{equation}\label{eq:P1-Unif-heated}
f_{\st}(\cancel{x},v)=\left(2\pi T_{\st}\right)^{-1/2} \exp\left[-\frac{v^{2}}{2T_{\st}}\right].
\end{equation}
With this ``stochastic thermostat'' forcing, the system remains
homogeneous for all times if it is initially so, as is also the case
of a inelastic gas of hard particles described by the inelastic
Boltzmann equation \cite{van_noije_velocity_1998}.

\section{Global stability}
\label{global-stability}

In this section, we analyse the global stability of the nonequilibrium
stationary solutions of the kinetic equation \eqref{eq:P1-hydroMM}
submitted to quite a general class of boundary conditions.  Following
the discussion in the introduction, we define the $H$-functional as
\begin{equation}\label{H-functional-st}
H[f]=\int\!\! dx\,dv f(x,v,t) \ln\!\left[\frac{f(x,v,t)}{f_{\st}(x,v)}\right].
\end{equation}

Let us consider the time evolution of $H[f]$. It is directly obtained
that 
\begin{align}
\frac{dH}{dt}=\int\!\! dx\, dv\, \tder{f}
\ln\left(\frac{f}{f_{\st}}\right)=\int\!\! dx\, dv\,\mathcal{L}f \,\ln\left(\frac{f}{f_{\st}}\right),
\label{eq:H-time-ev-1}
\end{align}
where $\mathcal{L}$ stands for the nonlinear evolution operator on the
rhs of the kinetic equation \eqref{eq:P1-hydroMM}, that is,
$\tder{f}=\mathcal{L}f$. Now we note the following property: if we
define $\Delta f=f-f_{\st}$ to be the deviation of the PDF from the
steady state, the linear terms in the deviation vanishes, since both
factors in the integrand of \eqref{eq:H-time-ev-1} are equal to zero
for $f=f_{\st}$. This is a desirable property: were it not true, the
sign of $dH/dt$ could be reversed for initial conditions close enough
to the steady state by simply reversing the initial value of
$\Delta f$. Thus, the existence of an $H$-theorem would be utterly
impossible, see also next Section.

Then, we can write
\begin{align}
\frac{dH}{dt}=\int\!\! dx\, dv\,\mathcal{L}f
  \,\ln\left(\frac{f}{f_{\st}}\right)-\int\!\! dx\,dv\, \mathcal{L}f_{\st}\, \frac{f-f_{\st}}{f_{\st}}.
\label{eq:H-time-ev-2}
\end{align}
Now, the idea is to split the operator $\mathcal{L}$ into the three
contributions on the rhs of Eq.~\eqref{eq:P1-hydroMM}: first, the
diffusive one; second, the one proportional to $\nu$, which is intrinsically dissipative; and third,
the one proportional to the noise strength $\xi$:
$\mathcal{L}_{\diff}$, $\mathcal{L}_{\inel}$ and
$\mathcal{L}_{\noise}$, respectively.  Accordingly, we have that the
time derivative of $H$  has
three contributions, 
\begin{equation}\label{eq:H-time-ev-total}
\frac{dH}{dt}=\left.\frac{dH}{dt}\right|_{\diff}+\left.\frac{dH}{dt}\right|_{\inel}+\left.\frac{dH}{dt}\right|_{\noise},
\end{equation}
obtained by inserting into Eq.~\eqref{eq:H-time-ev-2} the relevant
part of the evolution operator $\mathcal{L}$. Note that, although
$\mathcal{L}f_{\st}=0$, in general $\mathcal{L}_{\diff}f_{\st}\neq 0$,
$\mathcal{L}_{\inel}f_{\st}\neq 0$ and
$\mathcal{L}_{\noise}f_{\st}\neq 0$.

After some tedious but easy algebra, a summary of which is given
in Appendix \ref{sec:dHdt-general}, the following expressions are
derived. Firstly, for the diffusive term,
\begin{equation}\label{eq:H-time-ev-diff}
\left.\frac{dH}{dt}\right|_{\diff}=-\int\!\! dx\,dv\,f
\left(\xder{\ln f}-\xder{\ln f_{\st}}\right)^{2}\leq 0.
\end{equation}
Secondly, for the inelastic term, proportional to $\nu$, 
\begin{equation}\label{eq:H-time-ev-inel}
\left.\frac{dH}{dt}\right|_{\inel}=-\frac{\nu}{2}\int\!\! dx
\left(u-u_{\st}\right) \int\!\! dv
f \, \vder{\ln f_{\st}}.
\end{equation}
Finally, the noise term, proportional to $\xi$, reads
\begin{equation}\label{eq:H-time-ev-noise}
\left.\frac{dH}{dt}\right|_{\noise}=-\frac{\xi}{2}\int\!\! dx\,dv\,f
\left(\vder{\ln f}-\vder{\ln f_{\st}}\right)^{2}\leq 0.
\end{equation}
These results, and the following throughout this section, are valid
for a quite general set of boundary conditions, leading to the
cancellation of all the boundary terms arising after integrating by
parts, as detailed in Appendix~\ref{sec:dHdt-general}. This set
includes but is not limited to the Lees-Edwards and periodic boundary
conditions corresponding to the sheared and uniformly heated
situations, respectively. For instance, they also apply to the Couette
state, in which the system is driven by keeping its two edges at two
(in general, different) fixed temperatures $T_{L}$ and $T_{R}$.

The inelastic term $dH/dt|_{\inel}$ in Eq.~\eqref{eq:H-time-ev-inel}
does not have a definite sign in general. Therefore, it is the
inelastic term that prevents us from proving $H$ to be a
non-increasing function of time. It must be stressed that the
diffusive, inelastic and noise contributions to $dH/dt$ in
Eqs.~\eqref{eq:H-time-ev-diff}-\eqref{eq:H-time-ev-inel} come
exclusively from the diffusive, noise and inelastic contributions in
the kinetic equation, respectively, only once the linear terms has
been subtracted as is done in Eq.~\eqref{eq:H-time-ev-2}: see
Appendix~\ref{sec:dHdt-general} for details.

Despite the above discussion, global stability of the steady state can
be established without proving an $H$-theorem. The key point is the
following: the long time limit of $dH/dt$ is non-positive and thus $H$
has a finite limit, since it is bounded from below. Therefore, $dH/dt$
tends to zero in the long time limit and it can be shown that this is
only the case if
$f(x,v,\infty)\equiv\lim_{t\to\infty}f(x,v,t)=f_{\st}(x,v)$. 

The average velocity $u(x,t)$ satisfies a diffusive equation
\eqref{eq:hydroMMu}, and thus it irreversibly tends to the steady
profile corresponding to the given boundary conditions in the long
time limit. Therefore,
$u(x,\infty)\equiv\lim_{t\to\infty}u(x,t)=u_{\st}(x)$ and taking into
account Eq.~\eqref{eq:H-time-ev-inel},
\begin{equation}\label{eq:lim-long-time-dH/dt-2}
\lim_{t\to\infty}\left.\frac{dH}{dt}\right|_{\text{inel}}=0
\Rightarrow \lim_{t\to\infty}\frac{dH}{dt}\leq 0.
\end{equation}
Since $H[f]$ is bounded from below, the only possibility is
\begin{equation}\label{eq:lim-long-time-dH/dt}
\lim_{t\to\infty}\frac{dH}{dt}=0,
\end{equation}
and all the contributions to $dH/dt$ in
Eqs.~\eqref{eq:H-time-ev-diff}-\eqref{eq:H-time-ev-noise} vanish in
the long time limit. The vanishing of Eq.~\eqref{eq:H-time-ev-diff}
imposes that $f(x,v,\infty)=f_{\st}(x,v)\phi(v)$, where $\phi(v)$ is
an arbitrary function of $v$. For $\xi\neq 0$,
Eq.~\eqref{eq:H-time-ev-noise} implies that $\phi(v)$ must be a
constant, independent of $v$, and normalisation yields
$\phi(v)=1$. For $\xi=0$, Eq.~\eqref{eq:H-time-ev-noise} identically
vanishes but it can be also shown that $\phi(v)=1$ by using the
kinetic equation in the limit as $t\to\infty$. Therefore, for
arbitrary $\xi$, including $\xi=0$, we have that
\begin{equation}\label{eq:global-stable}
f(x,v,\infty)=f_{\st}(x,v).
\end{equation}
This completes the proof. The steady distribution $f_{\st}(x,v)$ is
globally stable: each time evolution $f(x,v,t)$ (corresponding to
a given initial condition) tends to it in the long time limit.

\section{Inadequacy of $H_{B}$ as a Lyapunov functional for
  non-conservative systems}\label{sec:inadequacy}

Here we show that Boltzmann's $H_{B}[f]$ cannot be used to build a
Lyapunov functional for intrinsically dissipative systems, in
agreement with the numerical results by Marconi et
al.~\cite{marconi_about_2013}. Not only do we prove it for the
simplified models considered here, but for a general kinetic equation
in which energy is not conserved in collisions, such as the inelastic
Boltzmann or Enskog equations. To keep the notation simple, we still
write $\tder{f}=\mathcal{L}f$, but now $\mathcal{L}$ stands for the
evolution operator in the considered kinetic description, which is
nonlinear in general.

First, we restrict ourselves to homogeneous
situations and thus drop the integral over $x$,
\begin{subequations}
\begin{align}
H_{B}[f]=&\int\!\! dv f \ln f \\
\frac{dH_{B}}{dt}=&\int\!\! dv\, \tder{f}
\ln f=\int\!\! dv\,\mathcal{L}f \,\ln f,
\label{eq:HB-time-ev-1}
\end{align} 
\end{subequations}
 Also, we consider a
system that is initially close to the steady state, such that we can
expand everything in powers of $\Delta f=f-f_{\st}$. Then,
\begin{equation}
\mathcal{L} f\equiv \mathcal{L}(f_{\st}+\Delta
f)=\cancelto{0}{\mathcal{L}f_{\st}}+\mathcal{L}_{\text{lin}}\Delta
f+O(\Delta f)^{2},
\end{equation}
in which $\mathcal{L}_{\text{lin}}$ is the linearised evolution
operator. Neglecting $O(\Delta f)^{2}$ terms, the linear approximation
arises,
\begin{align}
\left.\frac{dH_{B}}{dt}\right|_{\text{lin}}=\int\!\! dv\,
  (\mathcal{L}_{\text{lin}}\Delta f) \,\ln
  f_{\st}=\left.\frac{d}{dt}\la\ln f_{\st}\ra\right|_{\text{lin}}.
\label{eq:HB-time-ev-linear}
\end{align} 
On the one hand, the linear contribution vanishes in the elastic case:
$\ln f_{\st}$ is a sum of constants of motion, which are unchanged by
the linearised kinetic operator. Then, $H_{B}$ can be a candidate for
a Lyapunov functional. On the other hand, only mass and linear
momentum are conserved for non-conservative interactions.  Thus, no
longer is $\ln f_{\st}$ a sum of conserved quantities, and
\begin{align}
\left.\frac{dH_{B}}{dt}\right|_{\text{lin}}\neq 0.
\label{eq:HB-time-ev-linear-2}
\end{align} 
Therefore, by changing the initial sign of $\Delta f=f-f_{\st}$, which
can always be done, the initial sign of $dH_{B}/dt$ is reversed and
$H_{B}$ cannot be a Lyapunov functional. 

\begin{figure}
\centering
  \includegraphics[width=0.49 \textwidth]{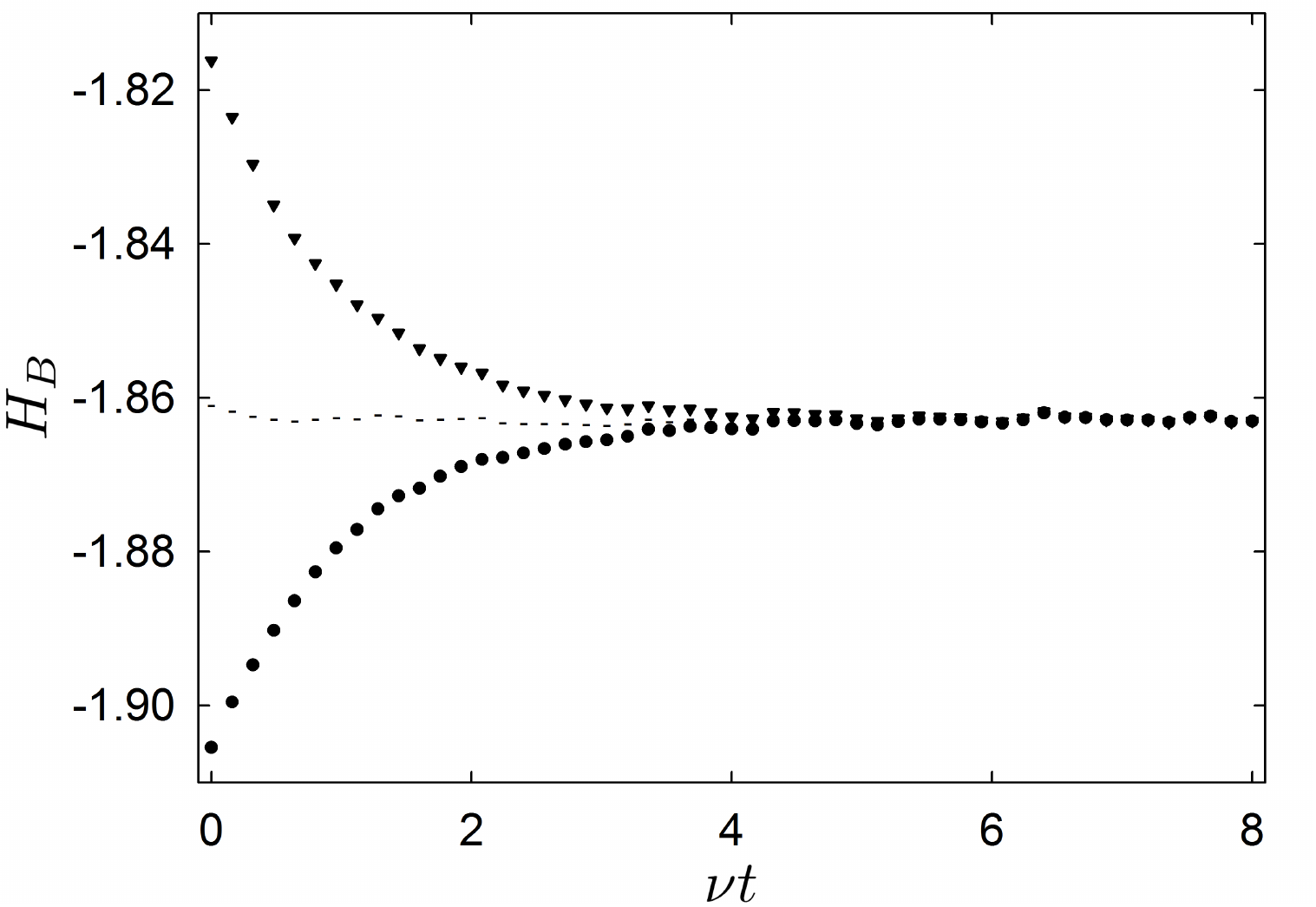}
  \caption{\label{fig:HBreversed} Evolution of the functional $H_{B}$
    for two different initial conditions in a uniformly heated system.
    Both simulations start from the stationary Gaussian shape but with
    a homogeneous temperature slightly shifted from the stationary
    one: (i) $T(t=0)=1.1 \, T_{\st}$ (circles) and (ii)
    $T(t=0)=0.9 \, T_{\st}$ (triangles). As predicted by the linear
    approximation, both functionals are symmetric with respect to the
    stationary value, which is schematically illustrated by plotting
    their mean value (dashed line). A system with $N=330$ sites has
    been considered, with $\nu=20$ and $\xi=50$. The plots correspond
    to averages over $3000$ trajectories.}
\end{figure}

In Fig.~\ref{fig:HBreversed}, we show the evolution of $H_{B}$ in our
kinetic model. We consider a uniformly heated system, so that the
system remain homogeneous for all times, as described in
Sec.~\ref{NESS}. Two different initial conditions are considered,
corresponding to Gaussian distributions with zero average velocity but
non-steady values of the temperature, specifically $1.1 \, T_{\st}$ and
$0.9 \, T_{\st}$. We can see how, in agreement with our discussion, not
only is one of the functionals increasing, but also it can be obtained
as the mirror image of the decreasing one through the stationary
value. Technical details about the simulation are provided in Appendix
\ref{sec:NumAp}.

Taking into account the specific (Gaussian) shape of the steady PDF
for the uniformly heated system, as given by
Eq.~\eqref{eq:P1-Unif-heated}, the time derivative of $H_{B}$ in
Eq.~\eqref{eq:HB-time-ev-linear} reduces to
\begin{equation}
  \label{eq:dHB/dt-Unif-Heated}
\frac{dH_{B}}{dt}=-\frac{1}{2T_{\st}}\frac{d\langle v^{2}\rangle}{dt}.
\end{equation}
Since the plots in Fig.~\ref{fig:HBreversed} correspond to evolutions
of the system for which $u(x,t)\equiv 0$ for all times, therein
$\langle v^{2}\rangle =T$ and, consistently, the $H_{B}$-curve
corresponding to an initial value of the temperature that is higher
(lower) than the steady one monotonically increases (decreases).

This is consistent with the situation found in
Ref.~\cite{bena_stationary_2006}, in which the uniformly heated
granular gas described by the inelastic Boltzmann equation was
investigated within the first Sonine approximation. Therein, the
entropy production was shown to have linear terms in the deviations of
the temperature and the excess kurtosis. Also, our result is
consistent with the numerical results in
Ref.~\cite{marconi_about_2013} for several collision models. The above
argument also proves why $H_{B}$ is not non-increasing for an elastic
system immersed in a heat bath at a temperature different from the
initial temperature of the gas, as also observed in
Ref.~\cite{marconi_about_2013}. Although $\ln f_{\st}$ is conserved in
collisions, the evolution operator includes a term coming from the
interaction with the bath that does not conserve the kinetic energy,
and again $dH/dt|_{\text{lin}}\neq 0$, making it impossible for
$H_{B}$ to be a Lyapunov functional.

In spatially non-homogeneous situations, the main difference is that
an additional integral over $x$ is present, both in $H_{B}$ and,
consequently, $dH_{B}/dt$. There is no reason to expect this integral
over space to make $dH/dt|_{\lin}$ vanish, since one still has that
\begin{equation}
\left.\frac{dH_{B}}{dt}\right|_{\lin}=\left.\frac{d}{dt}\la\ln f_{\st}\ra\right|_{\lin},
\end{equation}
and, in general, $\ln f_{\st}$ is not a sum of constants of motion. In
fact, again the sign of $\left.dH_{B}/dt\right|_{\lin}$ is reversed
when $\Delta f\to-\Delta f$, similarly to the homogeneous case. We
have numerically checked this prediction for the sheared system, with
the resulting evolution of $H_{B}$ being completely similar to that
for the uniformly heated case in Fig.~\ref{fig:HBreversed}, and thus
is not shown here.

\section{$H$-theorem for some specific NESS}
\label{USF}

Here we prove that the functional $H[f]$ is monotonically decreasing
for all times in some specific physical situations. Our proof applies
both to the sheared and uniformly heated systems described in
Sec.~\ref{NESS}. To be as general as possible, we consider a system
that is both heated and sheared: $a\neq 0$ and
$\xi\neq 0$.  In this situation, the boundary conditions for the
PDF are given by Eq.~\eqref{eq:P1-bc-USF}, which lead to
Eqs.~\eqref{eq:bc-USF-u} and \eqref{eq:bc-USF-T} for the averages
$u(x,t)$ and $T(x,t)$.

The steady solution of the hydrodynamic equations is
\begin{equation}\label{eq:USF-st-hydrovar}
u_{\st}(x)=a\left(x-\frac 1 2\right), \qquad T_{\st}=\frac{2a^{2} + \xi}{\nu}.
\end{equation}
On the one hand, the average velocity has a linear profile, similarly
to the situation in the USF state. On the other hand, the  temperature
remains homogeneous but its steady value has two contributions, one
coming from the shear and the other from the stochastic thermostat. The
viscous heating $2(\xder{u})^{2}$ and uniform heating $\xi$ terms
cancel the cooling term $-\nu T$ for all $x$.  The stationary solution
of the kinetic equation is readily found:
\begin{equation}\label{eq:Gauss-st-P1}
  f_{\st}(x,v)=(2\pi T_{\st})^{-1/2}\exp\left[-\frac{\left(v-u_{\st}(x)\right)^{2}}{2T_{\st}}\right],
\end{equation}
that is, the Gaussian distribution corresponding to the hydrodynamic
fields in Eq.~\eqref{eq:USF-st-hydrovar}.  Of course, the USF state
and NESS of the uniformly heated system in Sec.~\ref{NESS} can be
easily recovered as particular cases of Eq.~\eqref{eq:Gauss-st-P1}:
for $(a \neq 0,\xi=0)$ and $(a=0,\xi >0$), respectively.

Then, we turn now to the question of the existence of an $H$-theorem,
that is, the existence of a nonequilibrium entropy ensuring the
monotonic approach of the one-particle PDF to the steady state. Our
starting point is the following expansion of the one-particle PDF in
Hermite polynomials, 
\begin{align}
f(x,v,t)=&\frac{1}{\sqrt{2\pi
            T(x,t)}}\exp\left[-\frac{[v-u(x,t)]^{2}}{2T(x,t)}\right]\nonumber
            \\
& \times \left[1+\sum_{n=3}^{\infty}
\gamma_{n}(x,t)\,H_{n}\!\!\left(\frac{v-u(x,t)}{\sqrt{T(x,t)}}\right)\right],
\label{eq:USF-Hermite-expansion}
\end{align}
which is known as the Gram-Charlier series
\cite{Ch90,Ch05,Ed05,Wa58}. Therein, $u(x,t)$ and $T(x,t)$ are
the (exact) average velocity and temperature stemming from the
hydrodynamic equations for the considered distribution. The above
expansion is suggested by the Gaussian shape of the stationary PDF in
Eq.~\eqref{eq:Gauss-st-P1}. Now we define
\begin{align}
c=\frac{v-u(x,t)}{\sqrt{T(x,t)}}, \quad \tilde{f}(x,c,t)=\sqrt{T(x,t)}\,f(x,v,t),
\end{align}
From the orthogonality relation of the Hermite polynomials
\cite{AS72}, it is readily obtained that
\begin{equation}\label{eq:kappa-n}
\gamma_{n}(x,t)=\frac{1}{n!}\int\!\! dc\, H_{n}(c) \tilde{f}(x,c,t).
\end{equation}
Also, we could write $\gamma_n$ as a combination of moments of the
distribution. 

Some comments on the Gram-Charlier expansion are pertinent. First,
note that $n\geq 3$ in the sum: $\gamma_{1}=\gamma_{2}=0$ because the
zero-th order Gaussian contribution exactly gives the first two
moments $u(x,t)$ and $\la v^{2}\ra(x,t)=u^{2}(x,t)+T(x,t)$. Second, if
$f(x,v,t)$ were symmetric with respect to $v=u$, that is,
$\la(v-u)^{2n+1}\ra=0$ for all $n\in\mathbb{N}$, only even values of
$n$ would be present in the sum and one would end up with the usual
expansion in Sonine-Laguerre polynomials of kinetic theory.  Finally,
it is worth stressing that the series \eqref{eq:USF-Hermite-expansion}
converges for functions such that the tails of $\tilde{f}(x,c,t)$
approach zero faster than $e^{-c^2/4}$  for
$c\to\pm\infty$ \cite{Cr25,Sz39,Wa58}.

After a lengthy but straightforward calculation, which is summarised in
Appendix \ref{sec:dHdt<0-USF}, it is shown that 
\begin{equation}\label{eq:dH-dt-USF}
\frac{dH}{dt}=A(t)+B(t), \quad \text{with both } A(t),B(t)\leq 0.
\end{equation}
The expressions for $A(t)$ and $B(t)$ are
\begin{equation}\label{eq:USF-A(t)}
A(t)=-\int\!\! dx\, T \left[ \left( \frac{u^{\prime}}{T}-\frac{u_{\st}^{\prime}}{T_{\st}}\right)^{2}
+ \frac{\xi}{2} \left( \frac{1}{T}-\frac{1}{T_{\st}}\right)^{2} \right]
\end{equation}
and
\begin{widetext}
\begin{align}
B(t)=&-\frac{1}{\sqrt{2\pi}}\int\!\! dx\, dc \, \frac{e^{-c^{2}/2}}{1+\sum_{n=3}^{\infty} \gamma_{n} H_{n}(c)}
\left\{
  \frac{T^{\prime}}{2T}H_{2}(c) 
 +\sum_{n=3}^{\infty}\gamma_{n}^{\prime}H_{n}(c)-\sum_{n=3}^{\infty}\frac{\gamma_{n} u^{\prime}}{\sqrt{T}}n H_{n-1}(c) \right.
 \nonumber \\
 & \left.+\sum_{n=3}^{\infty}\frac{\gamma_{n}T^{\prime}}{2T}\left[H_{n+2}(c)+nH_{n}(c)\right] \right\}^2
 -\frac{\xi}{2\sqrt{2\pi}}\int\!\! dx\, dc \, \frac{e^{-c^{2}/2}}{1+\sum_{n=3}^{\infty} \gamma_{n} H_{n}(c)} \frac{1}{T}
\left[  \sum_{n=3}^{\infty} \gamma_{n} n H_{n-1}(c) \right]^2
 .
\label{eq:USF-B(t)}
\end{align}
\end{widetext}
We recall that the prime denotes spatial derivative.

Therefore, $dH/dt\leq 0$ for all times and we have shown that the
$H$-theorem holds for the sheared and heated system. Rigorously, our
proof holds for those PDFs such that the above Hermite expansion
converges. Note that the proof remains valid for the approach to any
NESS, whose PDF is a Gaussian with a homogeneous temperature,
independently of the corresponding boundary conditions.  In
Sec.~\ref{global-stability}, we have already demonstrated that $dH/dt$
only vanishes for $f(x,v,\infty)=f_{\st}(x,v)$, but the same result
can be rederived here in a different way. By imposing that both $A(t)$
and $B(t)$ vanish in the long time limit and making use of the
hydrodynamic equations for the averages, it can be shown that
$u(x,\infty)=u_{\st}(x)$, $T(x,\infty)=T_{\st}$ and
$\gamma_{n}(x,\infty)=0$, $\forall n\geq
3$.

\subsection{Numerical results for the USF state}

Here we consider the sheared system, and we numerically check our
theoretical predictions. Throughout this section, we use the values of
the parameters $\nu=20$, $a=5$ and $\xi=0$ (there is no stochastic forcing).

\begin{figure}
\centering
  \includegraphics[width=0.49 \textwidth]{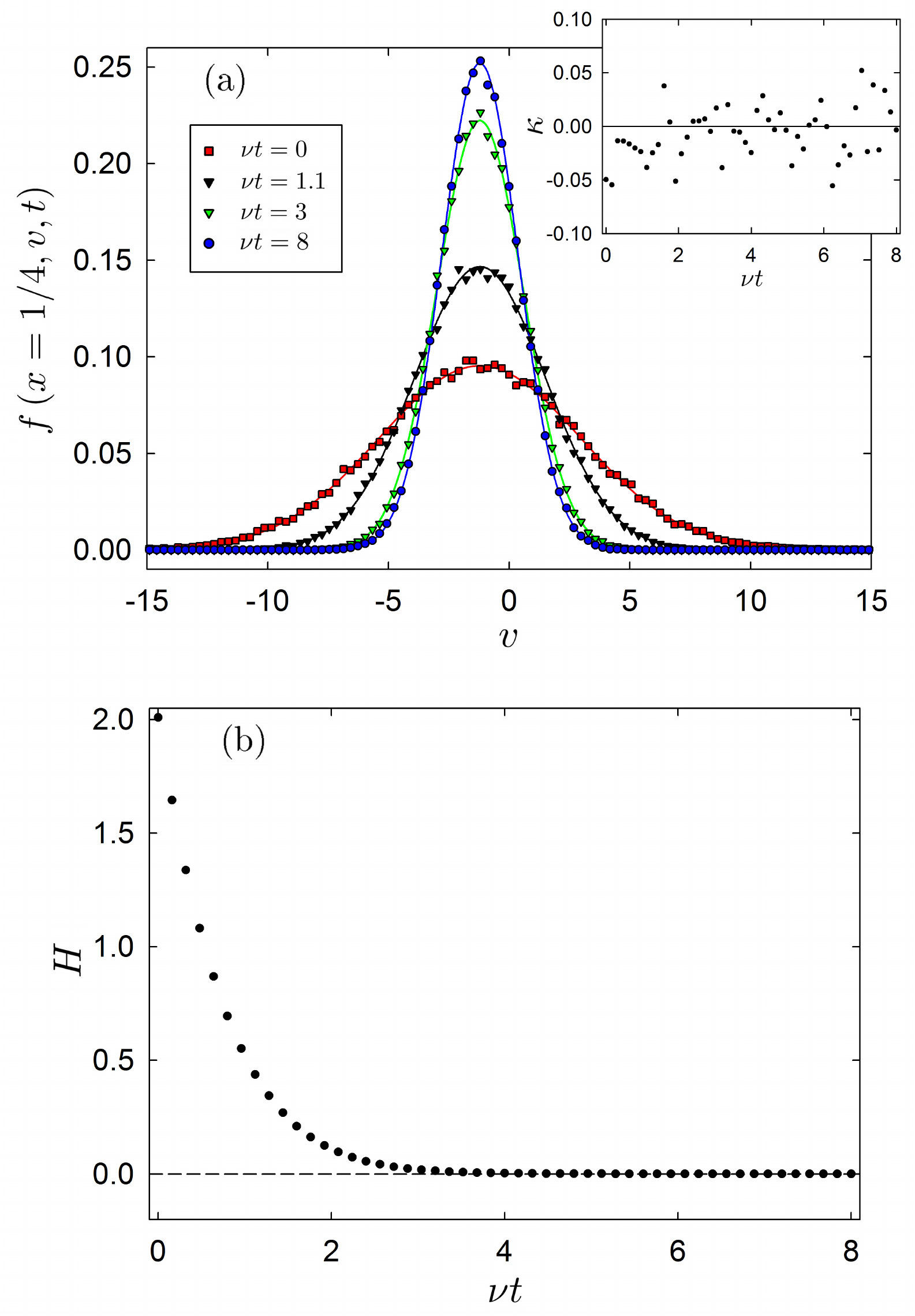}
  \caption{\label{fig:Gauss_hom} (Color online) Relaxation towards the USF state.
    The initial condition is Gaussian, centered in $u_{\st}(x)$ and
    with variance $T=7 \, T_{\st}$. (a) Velocity distribution
    function at $x=1/4$ for $4$ different times. In the inset, the
    evolution of the excess kurtosis is shown.  (b) Monotonic
    relaxation of the $H$ functional, in clear agreement with the
    proven $H$-theorem.  Data is averaged over $6000$ trajectories in
    a system with $N=660$ sites, $\nu=20$, and $a=5$. Solid lines
    correspond to the (theoretical) Gaussian distributions for the
    plotted times, except for the last time in which it represents the
    theoretical steady distribution. }
\end{figure}

Firstly, in Figure~\ref{fig:Gauss_hom}, we show the evolution of the
distribution and the $H$-functional from a Gaussian initial condition
with the steady velocity profile $u(x,0)=u_{\st}(x)$ but a higher
temperature, $T(t=0)=7 \, T_{\st}$. In panel (a), we depict the
velocity distribution at $x=1/4$ for several times. All of them are
Gaussian, which agrees with the theoretical prediction of the kinetic
equation: when the initial velocity profile coincides with the steady
one and only the temperature is perturbed, an initially Gaussian PDF
remains Gaussian for all times. Indeed, we can see in the inset how
the excess kurtosis
$\kappa = \langle [v-u(x)]^4 \rangle / \langle [v-u(x)]^2
\rangle^{2}-3$
only fluctuates around zero at the considered position $x=1/4$,
consistently with the Gaussian shape. In
panel (b), it is clearly observed that the $H$-functional is
monotonically decreasing with time.

\begin{figure}
\centering
  \includegraphics[width=0.49 \textwidth]{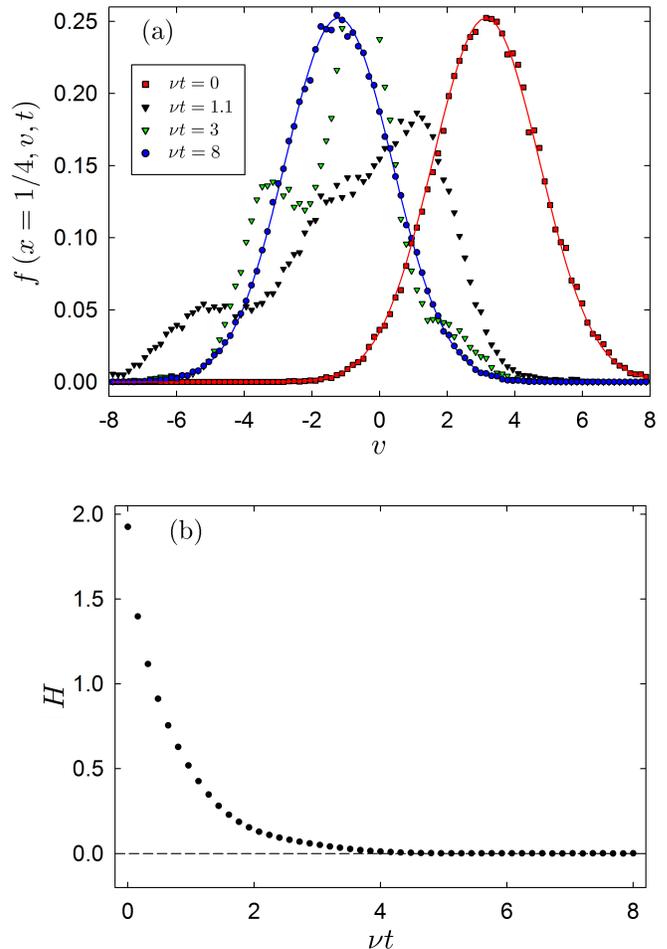}
  \caption{\label{fig:Gauss_Du} (Color online) The same plots as in
    Fig.~\ref{fig:Gauss_hom}, but starting from a different initial
    condition. Now, the initial PDF is a Gaussian centered in
    $u(x,0)=u_{\st}(x)+ 4.4 \, \sin (2 \pi x)$ and with variance
    $T(t=0)= T_{\st}$. In (a), solid lines correspond to
    the theoretical PDFs for the initial time and the steady state. In
    (b), $H$ decreases again monotonically towards its
    steady value, consistently with our theoretical prediction.}
\end{figure}

Secondly, we study the relaxation to the USF state from another
initial preparation, for which the velocity profile $u(x,0)$ is
different from the stationary but $T(x,0)=T_{\st}$. The numerical
results are shown in Fig.~\ref{fig:Gauss_Du}, and for the sake of
simplicity we use again an initial Gaussian
distribution. Specifically, we use
$u(x,0)=u_{\st}(x)+ 4.4 \, \sin (2 \pi x)$. Here, the departure from
the Gaussian shape is evident, and thus we have not plotted the
kurtosis. Consistently with our theoretical prediction, we get again a
monotonous relaxation of $H$ towards its null stationary value.

\begin{figure}
\centering
  \includegraphics[width=0.49 \textwidth]{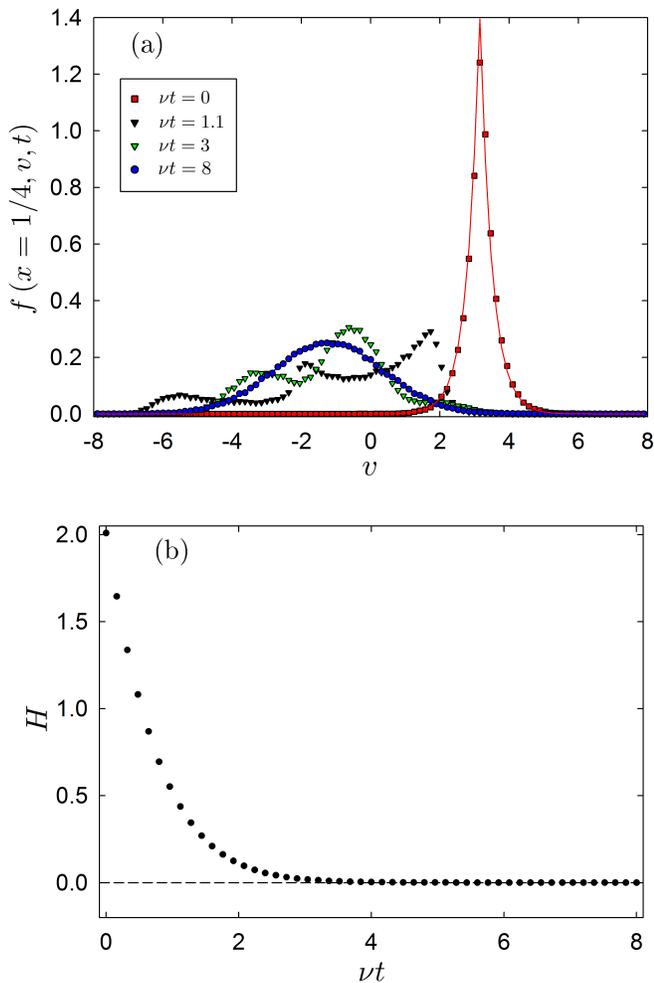}
  \caption{\label{fig:exp} (Color online) The same plots as in
    Fig.~\ref{fig:Gauss_hom}, but starting from an initial PDF with a
    divergent Gram-Charlier series. Concretely, the plots correspond
    to an exponential initial distribution centered in
    $u(x,t=0)=u_{\st}(x)+ 4.4 \, \sin (2 \pi x)$ and with
    $T(t=0)=0.1 \,T_{\st}$.  }
\end{figure}

Finally, we consider situations for which the above presented proof is
not rigorously applicable. As stated before, the Gram-Charlier series
does not converge when the tails of the distribution decay to zero
slower than the square root of the Gaussian. Nevertheless, when all
the coefficients $\gamma_{n}$ defined in Eq.~\eqref{eq:kappa-n} exist
and are finite, we still expect the $H$-theorem to hold. We illustrate
this situation with an initial exponential distribution; specifically,
we consider
\begin{equation}\label{eq:exp-PDF}
f(x,v,0) = \frac{1}{\sqrt{2 T(t=0)}} \exp \left[ \frac{ \sqrt{2} \left| v-u(x,t=0) \right| }{\sqrt{T(t=0)}} \right],
\end{equation}
with $u(x,t=0)= u_{\st}(x) + 4.4 \sin (2 \pi x)$ and
$T(t=0)=0.1 \,T_{\st}$.  Consistently with our expectation, we can see
in Fig.~\ref{fig:exp} that indeed the $H$-functional also
monotonically decreases.

\subsection{Numerical results in the uniformly heated system}

\begin{figure}
\centering
  \includegraphics[width=0.49 \textwidth]{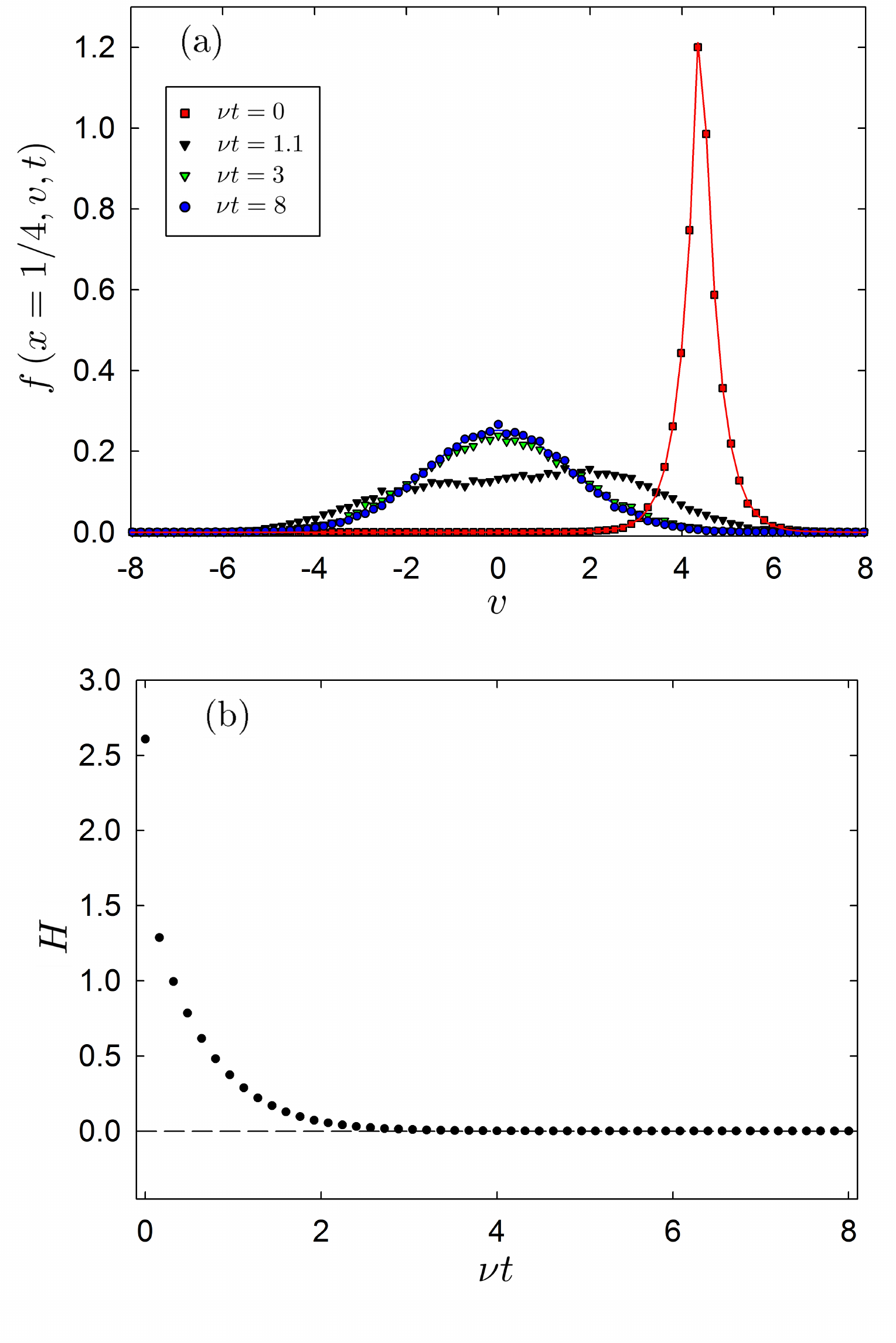}
  \caption{\label{fig:exp-heating} (Color online) Numerical results for the uniformly
    heated system.  The plots are analogous to those in
    Fig.~\ref{fig:Gauss_hom}, (a) time evolution
    of the PDF and (b) time evolution of the
    $H$-functional. The system is initially prepared with an
    exponential PDF centered in $u(x,t=0)= 4.4 \, \sin (2 \pi x)$ and
    with $T(t=0)=0.1 \,T_{\st}$.  Data is averaged over $3000$
    trajectories in a system with $N=330$ sites.  }
\end{figure}

To conclude, we put forward the results of simulations for the
uniformly heated system. Specifically, our simulations have been done
for $\nu=20$, $a=0$ (no shear) and $\xi=50$. In order not to overload
the reader with too many examples, we only present the more complex
case in Fig.~\ref{fig:exp-heating}: the relaxation towards the steady
state from an initial exponential distribution, as given by
Eq.~\eqref{eq:exp-PDF}. In particular, we consider that
$u(x,t=0)= 4.4 \, \sin (2 \pi x)$ and $T(t=0)=0.1 \, T_{\st}$. Note
that the perturbation from the steady values is the same as in
Fig.~\ref{fig:exp} for the sheared case. Again, we observe the
monotonic relaxation of $H$ towards the stationary value, consistently
with our theoretical result, even for a initial distribution for which
the Gram-Charlier series does not converge.

\section{Conclusions}
\label{conc}

Within a simplified model for a granular-like velocity field, we have
analytically shown that the nonequilibrium steady state that the
system eventually reaches in the long time limit is globally
stable. This has been done for quite a general situation, in which
energy may be injected into the system both through the boundaries and
by a heating mechanism that acts in the bulk. The proof
is valid both for spatially homogeneous situations (such as the
uniformly heated system) and inhomogeneous situations (such as the USF
or Couette states).

The proof of global stability is based on showing that the
$H$-functional $H=\int\! dx\, dv\, f \ln(f/f_{\st})$ (the
Kullback-Leibler divergence between the time evolving one-particle PDF
$f(x,v,t)$ and its value in the stationary state $f_{\st}(x,v)$
\cite{Kullback-Leibler}) is non-increasing in the infinite time limit. Thus,
we do not need $H$ to be a ``good'' Lyapunov functional for all times
in order to prove global stability. In conclusion, global stability
and the validity of an $H$-theorem do not seem to be unavoidably tied.

Moreover, we have analytically shown that the Boltzmann functional
$H_{B}=\int\! dx\, dv\, f \ln f$ cannot be, in general, a Lyapunov
functional for systems with non-conservative interactions. Close to
the steady state, we have proven that $dH_{B}/dt$ contains
non-vanishing terms that are linear in the deviations
$\Delta f=f-f_{\st}$. Therefore, a reversal of the sign of $\Delta f$
entails a reversal of the sign of $dH_{B}/dt$. This general analytical
proof on the inadequacy of $H_{B}$ as a Lyapunov functional is in
agreement with previous results in some specific cases
\cite{bena_stationary_2006,marconi_about_2013}.

We have also succeeded in demonstrating that the $H$-functional is
non-increasing and thus a ``good'' Lyapunov functional for some
specific driving mechanisms.  Our proof is not restricted to spatially
homogeneous situations and is applicable to two relevant physical
cases: the approach to (i) the USF state and (ii) the NESS
corresponding to the uniformly heated case.  The proof involves a
suitable expansion of the one-particle PDF in Hermite polynomials,
which is a generalisation of the well-known Sonine-Laguerre expansion
in kinetic theory. Although the proof is only rigorous for PDFs having
a convergent series expansion, we expect it to remain valid for more
general PDFs. In fact, we have numerically validated this expectation
in some specific situations.

The analytical results presented here are thus in agreement with
the numerical evidence in
Refs.~\cite{marconi_about_2013,de_soria_towards_2015} and advance the
understanding of this field in a twofold way. First, an analytical
proof, which was lacking, is provided for a simplified model.  Second,
spatially inhomogeneous situations are considered, both in the time
evolution and in its steady state.

Some limitations of our results have to be underlined, though. First,
the simplifications introduced in the model make it impossible to
address the problem of the stability of the homogeneous cooling state
in the undriven system at the level of the kinetic equation, as
already discussed in Ref.~\cite{manacorda_lattice_2016}. Second, for
the driving mechanisms for which we can analytically prove that $H$ is
a ``good'' Lyapunov functional, the steady distribution is exactly
Gaussian. Nevertheless, we think that this is not a fundamental point
and expect that the kind of expansion-based proof presented here may
be extended to other situations. A particularly appealing case is the
approach to the Couette NESS, for which the stationary PDF is
non-Gaussian in the model \cite{manacorda_lattice_2016}.

On a different note, our kinetic equation \eqref{eq:P1-hydroMM} shows
some resemblance to evolution equations for the one-particle PDF found
in other physical contexts, such as the Vlasov equation in plasma
physics or astrophysics
\cite{chavanis_statistical_1996,chavanis_jupiters_1998,yamaguchi_stability_2004}
or the non-linear (in the distribution function) Fokker-Planck
equation for systems of infinitely many coupled non-linear oscillators
exhibiting phase transitions \cite{desai_statistical_1978}. It is a
drift-like term depending on a certain average of the PDF that all
these different problems share. Both for the Vlasov and the non-linear
Fokker-Planck equations, the existence of a Lyapunov functional has
been proved by considering a variant of the functional $H[f]$ defined
in Eq.~\eqref{H-one-particle}
\cite{shiino_h-theorem_1985,shiino_dynamical_1987,bonilla_h-theorem_1996}. Thus,
an interesting prospect is to investigate if this kind of approach may
be extended to our class of models with non-conservative interactions.

Our work also opens the door to applying the ideas developed in this
paper to more complex models, closer to real non-conservative systems,
like granular gases. The pioneering numerical work in
Refs.~\cite{marconi_about_2013,de_soria_towards_2015} strongly
suggests that the $H$-functional is a ``good'' Lyapunov functional for
granular fluids. It seems worth trying to analytically prove that this
is indeed the case for the inelastic Boltzmann equation, at least for
some specific situations. If nothing else, one would like to be able
to show that the long time solutions are globally stable by showing
that $H$ is asymptotically non-increasing, similarly to what has been
done here.

\acknowledgments

We acknowledge the support of the Spanish Ministerio de Econom\'{\i}a y Competitividad through Grant FIS2014-53808-P. Carlos A. Plata also acknowledges the support from the FPU Fellowship Programme of the Spanish Ministerio de Educaci\'on, Cultura y Deporte through Grant FPU14/00241.

\appendix

\section{Derivation of the expression for $dH/dt$ in a general driven
  state}\label{sec:dHdt-general}

Let us consider the three contributions to $dH/dt$ in
Eq.~\eqref{eq:H-time-ev-total}. We start with the diffusive
one, 
\begin{align}\label{eq:dH/dt-diff-start}
\left.\frac{dH}{dt}\right|_{\diff}=\int\!\! dx\, dv\,\mathcal{L}_{\diff}f
  \,\ln\left(\frac{f}{f_{\st}}\right)-\int\!\! dx\,dv\,\frac{f}{f_{\st}}\, \mathcal{L}_{\diff}f_{\st} ,
\end{align}
where $\mathcal{L}_{\diff}f=\xxder{f}$ and we have used that
$\int\!\! dx\, dv\, \xxder{f_{\st}}$ vanishes identically. Integrating
by parts the first term on the rhs of Eq.~\eqref{eq:dH/dt-diff-start},
the result is
\begin{equation}
\int\!\! dv\, \xder{f}
\left.\ln\left(\frac{f}{f_{\st}}\right)\right|_{0}^{1}-\int\!\! dx\,
  dv\, f \xder{\ln f} \left( \xder{\ln f}-\xder{\ln f_{\st}}\right).
\end{equation}
Also integrating by parts the second term, one obtains
\begin{equation}
-\int\!\! dv\, \left.\frac{f}{f_{\st}} \xder{f_{\st}}\right|_{0}^{1}
+\int\!\! dx\,
  dv\, f \xder{\ln f_{\st}} \left( \xder{\ln f}-\xder{\ln f_{\st}}\right).
\end{equation}
We assume that the boundary terms are equal to zero, that is,
\begin{equation}
\int\!\! dv\,\left[\xder{f} \ln\left(\frac{f}{f_{\st}}\right)-\frac{f}{f_{\st}}
    \xder{f_{\st}}\right]_{0}^{1}=0.
\end{equation}
This is obviously true for Lees-Edwards and periodic boundary conditions \footnote{For the Couette state,
  in which the PDF at the boundaries is Gaussian with zero average
  velocity and a given temperature $T_{B}$ for all times, the first
  term is identically zero and the second vanishes because $\int\!\!
  dv\, f_{\st}(x,v)=1$ for all $x$}. Summing the two contributions to
the diffusive term above, we have 
\begin{align}\label{eq:dH/dt-diff-final}
\left.\frac{dH}{dt}\right|_{\diff}=-\int\!\! dx\, dv\, f \left( \xder{\ln f}-\xder{\ln f_{\st}}\right)^{2},
\end{align}
which is Eq.~\eqref{eq:H-time-ev-diff} of the main text. 

The noise term is treated along the same lines as above, but
integrating by parts in $v$ instead of $x$, since
$\mathcal{L}_{\noise}f=\frac{\xi}{2}\vvder{f}$. There in, the boundary
terms vanish if $f$ and $f_{\st}$ tend to zero fast
enough for $v\to\pm\infty$, and
\begin{align}\label{eq:dH/dt-noise-final}
\left.\frac{dH}{dt}\right|_{\noise}=- \frac{\xi}{2}\int\!\! dx\, dv\, f \left( \vder{\ln f}-\vder{\ln f_{\st}}\right)^{2},
\end{align}
which is Eq.~\eqref{eq:H-time-ev-noise}.

Now we focus on the inelastic contribution,
\begin{align}\label{eq:dH/dt-inel-start}
\left.\frac{dH}{dt}\right|_{\inel}=\int\!\! dx\, dv\,\mathcal{L}_{\inel}f
  \,\ln\left(\frac{f}{f_{\st}}\right)-\int\!\! dx\,dv\, \frac{f}{f_{\st}}\, \mathcal{L}_{\inel}f_{\st} ,
\end{align}
in which $\mathcal{L}_{\inel}f=\frac{\nu}{2}\vder{[(v-u)f]}$. Then,
\begin{align}
\left.\frac{dH}{dt}\right|_{\inel}=&\;\frac{\nu}{2}\int\!\! dx\, dv\, \vder{[(v-u)f]}
  \,\ln\left(\frac{f}{f_{\st}}\right)
\nonumber \\
&-\frac{\nu}{2}\int\!\! dx\,dv\, \vder{[(v-u_{\st})f_{\st}]}\, \frac{f-f_{\st}}{f_{\st}}.
\label{eq:dH/dt-inel-develop-2}
\end{align}
Again, integrating by parts in $v$ (here we do not write the boundary
terms at $v\to\pm\infty$), the first term on the rhs of
Eq.~\eqref{eq:dH/dt-inel-develop-2} is
\begin{align}
-\frac{\nu}{2}\int\!\! dx\, dv\, (v-u) f 
  \, \left(\vder{\ln f}-\vder{\ln f_{\st}}\right),
\label{eq:dH/dt-inel-develop-21}
\end{align}
whereas the second term gives
\begin{align}
\frac{\nu}{2}\int\!\! dx\, dv\, (v-u_{\st}) f
  \, \left(\vder{\ln f}-\vder{\ln f_{\st}}\right).
\label{eq:dH/dt-inel-develop-22}
\end{align}
Summing up these two contributions, and taking into account that both
$u$ and $u_{\st}$ do not depend on $v$,
\begin{align}
\left.\frac{dH}{dt}\right|_{\inel}=\frac{\nu}{2} \int\!\! dx \, (u-u_{\st}) \int\!\! dv\, f \left(\vder{\ln f}-\vder{\ln f_{\st}}\right)
\label{eq:dH/dt-inel-final}
\end{align}
Since $\int\! dv\, f \,\vder{\ln f}\equiv \int\! dv\, \vder{f}=0$,
this leads to Eq.~\eqref{eq:H-time-ev-inel}.

\section{Simulation strategy}\label{sec:NumAp}
In the simulations, so as to generate a trajectory of the stochastic
process, we proceed as follows. (i) A pair $(l,l+1)$ is chosen at
random and undergoes the inelastic collision described by
Eq.~\eqref{eq:coll_rule}, (ii) all the particles are submitted to the
stochastic thermostat according to \eqref{jump-moments-1} and
\eqref{jump-moments-2}, and (iii) time is incremented by
$\delta\tau=-(N\omega)^{-1}\ln x$, with $x$ being a homogeneously
distributed random number in $(0,1)$
\cite{bortz_new_1975,GILLESPIE1976403,prados_dynamical_1997,serebrinsky_physical_2011}. This
cycle (random choice of a pair and noise interaction followed by a
time increment) is repeated until time exceeds some maximum time
$t_{\text{max}}$.

Regarding the measurements of $f(x,v,t)$, we sample both position and
velocity spaces by defining $N_x$ bins of width $\Delta x$ and $N_v$
bins of with $\Delta v$. Of course, the product $N_x \Delta x = 1$,
the whole lattice, whereas $N_v \Delta v$ gives the range of
velocities bounded by the cutoffs $v_\text{\text{min}}$ and
$v_{\text{max}}$. In our simulations, we control that the contribution
to the PDFs coming from velocities outside the considered interval
$[v_{\text{min}},v_{\text{max}}]$ is negligible. With such a binning,
we build up an histogram and therefrom the distribution function
$f(x,v,t)$, which is represented by a $N_x \times N_v $ matrix for
each time $t$. Both $H$ and $H_{B}$ are computed by numerically
replacing the integral over $x$ and $v$ with sums over the prescribed
bins.

\section{Derivation of Eq.~\eqref{eq:dH-dt-USF}}
\label{sec:dHdt<0-USF}

In order to derive Eq.~\eqref{eq:dH-dt-USF}, we have to substitute the
Gaussian stationary solution \eqref{eq:Gauss-st-P1} and the Gram-Charlier series \eqref{eq:USF-Hermite-expansion} into the three
contributions to $dH/dt$, given by Eqs.~\eqref{eq:H-time-ev-diff},
\eqref{eq:H-time-ev-inel} and \eqref{eq:H-time-ev-noise}. 

For the inelastic term, it is readily obtained that
\begin{equation}\label{eq:H-terms-1}
\left.\frac{dH}{dt}\right|_{\inel} = \frac{\nu}{2 T_{\st}}\int\!\! dx
\left(u-u_{\st}\right)^2.
\end{equation}
For the diffusive and noise terms, the key ideas are a changing the
integration over velocities from $v$ to $c=(v-u)/\sqrt{T}$ and the use
of the recursion relations and the orthogonality property of the
Hermite polynomials \cite{AS72}. Working along
these guidelines, we arrive at
\begin{eqnarray}\label{eq:H-terms-2}
\left.\frac{dH}{dt}\right|_{\diff} &=& -\int\!\! dx\, T \left( \frac{u^{\prime}}{T}-\frac{u_{\st}^{\prime}}{T_{\st}}\right)^{2} \nonumber 
\\
&& -\frac{u_{\st}^{\prime 2}}{T_{\st}^2}\int\!\! dx
\left(u-u_{\st}\right)^2 + B_{1}(t),
\\
\left.\frac{dH}{dt}\right|_{\noise} &=& - \frac{\xi}{2}\int\!\! dx\, T \left( \frac{1}{T}-\frac{1}{T_{\st}}\right)^{2} \nonumber 
\\
&& -\frac{\xi}{2T_{\st}^2}\int\!\! dx
\left(u-u_{\st}\right)^2 + B_{2}(t),
\end{eqnarray}
where $B_1$ and $B_2$ are, respectively, the first and the second term
in Eq.~\eqref{eq:USF-B(t)}. The sum of the factors multiplying
$\int\!\! dx \left(u-u_{\st}\right)^2$ vanishes by taking into account
the equation for the (spatially homogeneous) stationary
temperature. Therefore, the sum of the remaining terms leads right to
Eq.~\eqref{eq:dH-dt-USF}.

\bibliography{Mi-biblioteca-9-dic-2016-editado-por-Carlos-v2,Granular-v2,Series}

\end{document}